\documentclass{aa}

\input epsf 

\begin{document}

\sloppy

\thesaurus{08(13.07.2; 09.09.1 Crab Nebula; 09.19.2)} 

\title{Optimizing the angular resolution of the HEGRA telescope
system to study the emission
region of VHE gamma rays in the Crab Nebula}

\titlerunning{Study of the VHE emission region in the Crab Nebula}
\authorrunning{F. Aharonian et al.}

\author{F.A. Aharonian\inst{1},
A.G.~Akhperjanian\inst{7},
J.A.~Barrio\inst{2,3},
K.~Bernl\"ohr\inst{1},
H.~Bojahr\inst{6},
O.~Bolz\inst{1},
H.~B\"orst\inst{5},
J.L.~Contreras\inst{3},
J.~Cortina\inst{3},
S.~Denninghoff\inst{2}
V.~Fonseca\inst{3},
J.C.~Gonzalez\inst{3},
N.~G\"otting\inst{4},
G.~Heinzelmann\inst{4},
G.~Hermann\inst{1},
A.~Heusler\inst{1},
W.~Hofmann\inst{1},
D.~Horns\inst{4},
A.~Ibarra\inst{3},
C.~Iserlohe\inst{6},
I.~Jung\inst{1},
R.~Kankanyan\inst{1},
M.~Kestel\inst{2},
J.~Kettler\inst{1},
A.~Kohnle\inst{1},
A.~Konopelko\inst{1},
H.~Kornmeyer\inst{2},
D.~Kranich\inst{2},
H.~Krawczynski\inst{1},
H.~Lampeitl\inst{1},
E.~Lorenz\inst{2},
F.~Lucarelli\inst{3},
N.~Magnussen\inst{6},
O.~Mang\inst{5},
H.~Meyer\inst{6},
R.~Mirzoyan\inst{2},
A.~Moralejo\inst{3},
L.~Padilla\inst{3},
M.~Panter\inst{1},
R.~Plaga\inst{2},
A.~Plyasheshnikov\inst{1,}$^\S$,
J.~Prahl\inst{4},
G.~P\"uhlhofer\inst{1},
W.~Rhode\inst{6},
A.~R\"ohring\inst{4},
G.P.~Rowell\inst{1},
V.~Sahakian\inst{7},
M.~Samorski\inst{5},
M.~Schilling\inst{5},
F.~Schr\"oder\inst{6},
M.~Siems\inst{5},
W.~Stamm\inst{5},
M.~Tluczykont\inst{4},
H.J.~V\"olk\inst{1},
C.~Wiedner\inst{1},
W.~Wittek\inst{2}}

\institute{Max Planck Institut f\"ur Kernphysik,
Postfach 103980, D-69029 Heidelberg, Germany \and
Max Planck Institut f\"ur Physik, F\"ohringer Ring
6, D-80805 M\"unchen, Germany \and
Universidad Complutense, Facultad de Ciencias
F\'{i}sicas, Ciudad Universitaria, E-28040 Madrid, Spain 
\and
Universit\"at Hamburg, II. Institut f\"ur
Experimentalphysik, Luruper Chaussee 149,
D-22761 Hamburg, Germany \and
Universit\"at Kiel, Institut f\"ur Experimentelle und Angewandte Physik,
Leibnizstra{\ss}e 15-19, D-24118 Kiel, Germany\and
Universit\"at Wuppertal, Fachbereich Physik,
Gau{\ss}str.20, D-42097 Wuppertal, Germany \and
Yerevan Physics Institute, Alikhanian Br. 2, 375036 Yerevan, 
Armenia\\
\hspace*{-4.04mm} $^\S\,$ On leave from  
Altai State University, Dimitrov Street 66, 656099 Barnaul, Russia\\
}

\mail{Werner Hofmann, \\Tel.: (Germany) +6221 516 330,\\
email address: Werner.Hofmann@mpi-hd.mpg.de}

\offprints{Werner Hofmann}

\date{Received ; accepted }

\maketitle

\begin{abstract}
The HEGRA system of imaging atmospheric Cherenkov telescopes
provides for specially selected classes of events an angular
resolution of better than 3'. By comparing the measured angular 
distribution of TeV gamma rays from the Crab Nebula
with the distribution expected on the basis of Monte Carlo
simulations, and with measurements of gamma rays from the
point source Mrk 501, we conclude that the rms size of the VHE
gamma-ray emission region in the Crab Nebula is less than
1.5'. 


\end{abstract}

\section{Introduction}

{Stereoscopic systems of imaging atmospheric Cherenkov telescopes (IACTs)
such as the HEGRA IACT system (Daum et al. \cite{hegra_perf}) allow to
reconstruct the directions and energies of TeV gamma-rays with high
precision. The analysis techniques, 
the control of systematics, 
and
the understanding of the angular resolution function
of the instrument
(P\"uhlhofer et al. \cite{hegra_pointing},
Aharonian et al. \cite{hegra_501},
Hofmann et al. \cite{hegra_reco})
has progressed to a level that one can start to study the
characteristics of extended sources on a scale of a few arcminutes.
These scales start to become interesting in disentangling the 
emission mechanisms for Galactic TeV gamma-ray sources such as 
the Crab Nebula or the pulsar PSR B1706-44.
In this paper, we present, as a case study, an investigation of the
size of the emission region of the Crab Nebula at TeV energies.}

The Crab Nebula is one of the best-studied objects
in the sky, in all wavelength regimes. It has been 
established as a TeV gamma-ray source by the Whipple group,
using the imaging atmospheric Cherenkov technique
(Weekes et al. \cite{weekes}, Vacanti et al. \cite{vacanti}), 
and has been studied with
many other Cherenkov telescopes.
The precise spectral shape of gamma-ray emission from the Crab
Nebula has been the subject of a number of recent 
publications
(Hillas et al. \cite{whipple_c}, Tanimori et al. \cite{cangaroo}, 
Konopelko \cite{hegra_crab}). 
The spectrum is consistent with a power-law
extending from a few 100 GeV out to energies of 50 TeV
and beyond. Contrary to observations in the X-ray and
GeV gamma-ray regimes, the TeV gamma-ray emission does
not show a pulsed component attributable to a direct
contribution from the Crab Pulsar; pulsed emission is
below 3\% of the DC flux (Aharonian et al. \cite{hegra_pulsar},
Burdett et al. \cite{whipple_pulsar}). 
The commonly accepted model
for VHE gamma-ray production in the Crab Nebula assumes
electron acceleration in the termination shock of the
pulsar
wind at a distance of about 0.1 pc (0.2') from the pulsar
(see, e.g., Kennel \& Coroniti (\cite{coroniti_crab}), 
De Jager \& Harding (\cite{harding_crab}),
Atoyan \& Aharonian (\cite{aharonian_crab}),
Aharonian \& Atoyan (\cite{review})).
The electrons diffuse out into the Nebula and 
produce a characteristic two-component
electromagnetic spectrum: synchrotron emission dominates at most
energies {up to about 0.1 GeV}, 
whereas the inverse Compton
process generates higher-energy gamma-rays with energies
 from the GeV range up to 100 TeV and beyond. From the
relative strength of the two components, values for the
average magnetic field of 15 - 30 nT have been derived
(Hillas et al. \cite{whipple_c},
De Jager \& Harding \cite{harding_crab},
Atoyan \& Aharonian \cite{aharonian_crab}).

The Crab Nebula represents an extended source of
electromagnetic radiation. Since the electrons loose
energy as they expand out into the Nebula, primarily 
due to synchrotron losses, the effective source size 
is predicted to shrink with increasing energy of the
radiation, with radio emission extending up to and
beyond the filaments visible in the optical, whereas
hard X-rays and multi-TeV gamma-rays should be produced 
primarily in the direct vicinity of the shock
(see, e.g.,
Kennel \& Coroniti (\cite{coroniti_crab}),
De Jager \& Harding (\cite{harding_crab}),
Atoyan \& Aharonian (\cite{aharonian_crab}),
Amato et al. (\cite{amato})).
At TeV energies, a second production mechanism for
gamma-rays could be the hadronic production by
protons accelerated in the shock 
(Atoyan \& Aharonian \cite{aharonian_crab})
or resulting from decays of secondary
neutrons produced in the pulsar magnetosphere
(Bednarek \& Protheroe \cite{bednarek}); gamma rays are 
produced in interactions with the surrounding material,
e.g. in the filaments
(Atoyan \& Aharonian \cite{aharonian_crab}). This contribution might be
enhanced due to a trapping of protons in local magnetic
fields associated with the filaments, increasing  the 
interaction probability. Given that the size of the
Crab Nebula -- with its about 4' by 3' extension in the
optical -- is comparable to the angular resolution
achieved for TeV gamma rays by the HEGRA system of
imaging atmospheric Cherenkov telescopes (IACTs),
a study of the size of
the TeV emission region of the Crab Nebula is 
now  possible with meaningful sensitivity.
This paper reports such an analysis, based on data
collected over the last years with the HEGRA IACT 
system.

\begin{figure}
\begin{center}
\mbox{
\epsfxsize9.0cm
\epsffile{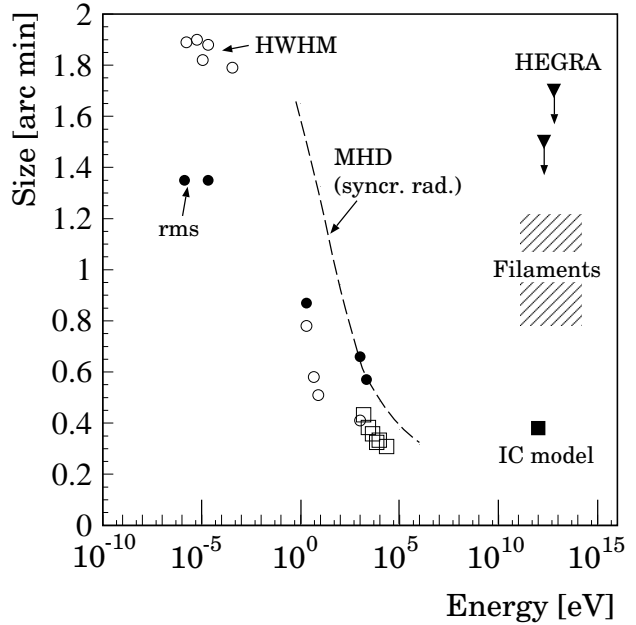}}
\label{crab_size}
\caption{Angular size of the Crab nebula at different frequencies.
Full circles: rms size, averaged over directions. Open circles:
half width at half maximum (HWHM, defined as FWHM/2), 
averaged over the long and short axis.
Open squares: HWHM along the $300^\circ$ direction. See text
for references. 
The dashed 
line indicates the frequency dependence of the size of the 
(synchrotron-radiation) emission
region as given by Kennel \& Coroniti (\cite{coroniti_crab}), and 
the full square shows the rms
size predicted for inverse-Compton gamma-rays at TeV energy 
(Atoyan \& Aharonian \cite{aharonian_crab}). 
The dashed region indicates the rms size range of the filaments,
the likely scale for hadronic production mechanisms. The triangles
show the upper limits on the rms source size at TeV energies derived
in this work.}
\end{center}
\end{figure}

For comparison and later reference, we will briefly
summarize the existing information on the size of
the Crab Nebula, as a function of the energy of
the radiation. An obvious problem in such a compilation
is that there is no unique definition of `size'.
For comparison with the TeV results given later,
the most relevant quantity is an rms size, gained
by approximating the intensity distribution by a 
two-dimensional Gaussian, or by directly calculating
the rms width by projecting or slicing the intensity
distribution along an axis, and averaging over 
directions. Rms width values based on 5 GHz radio data
(Wilson \cite{wilson}), optical data from 
Woltjer (\cite{woltjer}) as displayed
in Wilson (\cite{wilson}), and 0.1 - 4.5 keV X-ray data 
(Harnden \& Seward \cite{harnden}) are
compiled in Hillas et al. (\cite{whipple_c}). Additional rms values were
obtained for the 327 MHz radio data of 
Bietenholz et al. (\cite{bietenholz}), and for the 22 - 64 keV X-ray data of
Makishima et al. (\cite{makishima}). These data are shown in Fig. \ref{crab_size}
as full circles.
The bulk of the size values quoted in the literature
refer to a different measure, the full width
at half maximum (FWHM), which unfortunately in case of a
structured intensity distribution depends also on 
the resolution of the instrument. In some cases,
data are only available along a specific direction,
and do not allow to average over the long and short
axis of the Nebula. In particular in earlier X-ray
data, the contributions of the pulsar and of the
surrounding nebula are not separated. Fig. \ref{crab_size}
includes (as open circles) FWHM size data, averaged over the long and short axis,
at radio wavelengths (Wilson \cite{wilson}) and in the optical, NUV, FUV, and X-ray,
as given in Hennessy et al. (\cite{hennessy}). Also shown are X-ray data compiled in
Ku et al. (\cite{ku}), which refer to the width along the $300^\circ$ direction
(open squares). A clear trend for decreasing source size with 
increasing energy is evident, considering either the rms size values, the 
averaged FWHM size values, and the fixed-direction X-ray widths. 
Included as dashed line is the frequency dependence of the 
synchrotron emission region as sketched in 
Kennel \& Coroniti (\cite{coroniti_crab}). 
The size of the emission region for inverse-Compton TeV gamma
rays can be predicted using the average magnetic field to relate
synchrotron photon energies to electron energies and to
inverse-Compton gamma rays; from such arguments, one concludes
that the size for TeV gamma-rays should correspond to the X-ray size.
The rms
size predicted for the TeV gamma-ray emission region by the detailed 
calculations of 
Atoyan \& Aharonian (\cite{aharonian_crab})
 is included in Fig.~\ref{crab_size}. Basically,
inverse-Compton TeV gamma-rays should emerge from the toroidal
X-ray emission region clearly visible in the ROSAT data
(Hester et al. \cite{hester})
{and in the recent Chandra image (Weisskopf et al., 2000), 
as already speculated 
earlier by Aschenbach \& Brinkmann (\cite{aschenbach}).
The projected semi-major and semi-minor
axes of the emission torus are 38'' and 18'', respectively.
Due to the nonuniform strength of X-ray emission 
along the torus, the resulting emission profile is 
roughly elliptical, and its center is
shifted N relative to the pulsar location by about 0.3'.}
Hadronic production mechanisms are expected to generate larger source
sizes, of the scale of the size of the nebula (shaded region
in Fig.~\ref{crab_size}).

\section{Observations of the Crab Nebula with the HEGRA CT system}

The HEGRA system of imaging atmospheric Cherenkov telescopes
is located on the Canary Island of La Palma, on the
site of the Observatorio del Roque de los Muchachos. The
telecope system consists of five telescopes, with a
mirror area of 8.5 m$^2$ and a focal length of 5 m. 
A sixth prototype telescope is operated in stand-alone
mode and is not used for the analysis presented here.
The system telescopes are arranged at the  corners and in the
center of a square of about 100 m side length. The
alt-azimuth mounted telescopes are equipped with cameras
consisting of 271 photomultipliers (PMTs). Each PMT views an area
of the sky of $0.25^\circ$ diameter; the field of view of 
each camera is about $4.3^\circ$. Cherenkov images of
air showers are recorded whenever two telescopes trigger
simultaneously; the trigger condition requires that two neighboring 
PMTs exhibit signals equivalent to 8 or more photoelectrons.
Typical trigger rates are around 15 Hz, for an energy
threshold of 500 GeV for vertical gamma rays. Details
about the HEGRA IACTs and their performance can be found in
Daum et al. (\cite{hegra_perf}),
Aharonian et al. (\cite{hegra_501}),
Hermann (\cite{hegra_camera}),
Bulian et al. (\cite{hegra_trigger}).

The Crab Nebula was observed in each season since the HEGRA
IACT system commenced operation in late 1996, initially
with three telescopes, later with four and since late 1998
with the complete set of five telescopes. For this analysis,
only data taken in the years 1997 and 1998 were used,
acquired with at least four telescopes under good weather
conditions; in order to be able to compare with earlier Mrk 501
data taken with four telescopes, data from the fifth telescope 
was not used in the most recent five-telescope data sets.
 The quality-selected data set amounts to an
integral observation time of about 155 h, and includes about
6.3 million events. For the final analysis, only data
taken at zenith angles of less than $30^\circ$ were included,
with about 3.5 million events remaining.

The Crab Nebula was observed in the so-called wobble mode,
with the source offset by $0.5^\circ$ in declination relative to the
telescope axes. The sign of the offset alternated every 
20 min. A region offset by the same amount, but in the
opposite direction, is used for background estimates,
avoiding the need for special off-source observations. Since 
the stereoscopic reconstruction of air showers provides an
angular resolution of typically 6', signal and
background regions are well separated. 

The techniques for data analysis are similar to those
documented, e.g., in Aharonian et al. (\cite{hegra_501},\cite{hegra_501b}).
Direction and impact point of an air shower are
reconstructed from the stereoscopic views
of the shower. Based on the measured impact point
and the known (Aharonian et al. \cite{hegra_pool}) distribution of 
Cherenkov light as a function of the distance to the
shower axis, an energy estimate is derived, with a typical
resolution of 20\%. Gamma-ray candidates are selected
on the basis of image shapes. Given the core distance
and the intensity (the Hillas {\em size} parameter
(Hillas \cite{hillas_param})), the expected {\em width} of
a gamma ray image is determined. The measured
{\em width} values are normalized to this value, and
averaged over telescopes. A cut on the resulting
{\em mean scaled width} $<$ 1.2 retains most gamma-rays,
but rejects the bulk of the cosmic-ray showers.

In detail, the reconstruction of shower geometry
differs somewhat from the techniques used so far.
Whereas the normal reconstruction procedure combines images
from all telescopes regardless of their quality,
the new procedure assigns -- on the basis of
the Monte Carlo simulations of Konopelko et al. (\cite{hegra_mc}) --
errors to the relevant image parameters (the
location of the image centroid and the orientation
of the image axis). These errors depend on the
intensity and the shape of the images and are 
propagated through the geometrical reconstruction,
resulting in error estimates (or, to be precise,
a covariance matrix) for the shower parameters.
Details of the algorithm are given in 
Hofmann et al. (\cite{hegra_reco}). Depending on the characteristics
of an event, an angular resolution between 
2' and more then 10' is predicted,
with average value slightly  below 6'
\footnote{Here and in the following, ``angular
resolution'' is defined as the Gaussian width $\sigma$ of
the distribution of reconstructed shower directions,
projected onto one axis of a local coordinate system.}.
The ability to select subsets of events with
better-than-average resolution will be used
extensively in the analysis of the size of the
VHE emission region in the Crab Nebula.

\section{The limit on the size of the emission region}

Even if the size at radio wavelengths -- about 1.3' rms --
is used as a most extreme possibility for the size at
TeV energies, the source size $\sigma_s$ is still smaller than the
angular resolution $\sigma_o$ for the best subsets - about 2' to 3'. 
Therefore, one cannot expect to generate a detailed map of the
source.
Instead, an extended emission region of (rms) size $\sigma_s$
would primarily show up as
a slight broadening of the angular distribution of 
gamma-rays, beyond the value determined by the 
experimental resolution:
\begin{equation}
\label{eqa}
\sigma = \sqrt{\sigma_o^2+\sigma_s^2}
\end{equation}
or, for $\sigma_s$ small compared to $\sigma_o$,
\begin{equation}
\label{eqb}
{\Delta \sigma \over \sigma_o} =
{\sigma-\sigma_o \over \sigma_o} \approx 
{\sigma_s^2 \over 2 \sigma_o^2}
\end{equation}
For an intrinsic resolution
of 3' (in a projection)
and a 1.5' rms source size, one would find a 
3.4' wide angular distribution; for the 6' resolution,
the resulting width is 6.2'. In order to positively
detect a finite source size, or to derive stringent 
upper limits, one has (a) to measure the width of
the angular distribution of gamma-rays with sufficient
statistical precision, and (b) to quantitatively understand
the response function of the instrument at the same level,
and to control systematic effects which influence the
resolution. 

For a given number $n$ of events from the source, and
ignoring for the moment the effect of background under the
signal, the statistical error on the
width of the (projected) angular distribution is
$\Delta \sigma / \sigma = 1/(2 n^{1/2})$. In case the
angular distribution is consistent with a point source,
Eqn. \ref{eqb} then implies a (1 standard deviation) upper
limit $\sigma_s < \sigma_o n^{-1/4}$. Therefore, one will
want to minimize $\sigma_o$ by selecting a subset of events
with particularly well-determined directions, even at the expense of event
statistics. More critical are, in general, systematic errors.
Pointing imperfections, changes in mirror alignment, etc.
can cause differences in the angular resolution between
data sets taken at different times, under different conditions,
and between the data and the simulation. If, e.g., the intrinsic
resolution of the instrument is known and reproducible to 10\%, the 
minimum source size which can be reliably detected is
$ 0.46 \sigma_o$, according to Eqs. \ref{eqa} or \ref{eqb}. 
Once more, a selection towards small $\sigma_o$ is preferred.
Among gamma-ray events, we find that about 1\% of the events have
a predicted resolution below 1.8' ($0.03^\circ$), 6\% below 
2.4' ($0.04^\circ$), 15\% below 3' ($0.05^\circ$), and
60\% below 6' ($0.1^\circ$), respectively. In particular
the samples with resolutions better than 2.4' or
3' combine good angular resolution with
acceptable statistics. The cuts on angular resolution
have the additional benefit on enhancing the 
gamma-ray sample relative to the cosmic-ray background;
cosmic-ray showers generate more diffuse images
and have a worse angular resolution. Fig. 2
shows the angular distribution of events retained
after a cut at 3' resolution in both projections,
applying only very loose additional cuts on event
shapes (a cut on the {\em mean scaled width} at
1.2, which retains over 80\% of the gamma-ray events).
The selection also biases the sample towards higher
energies, since high-energy events produce more intense
images, with smaller errors on the image parameters.
In the overall data sample, the median energy of
reconstructed events is 0.9 TeV; after a cut on the
resolution at 3', this value rises to 2.0 TeV.
\begin{figure}
\begin{center}
\mbox{
\epsfxsize8cm
\epsffile{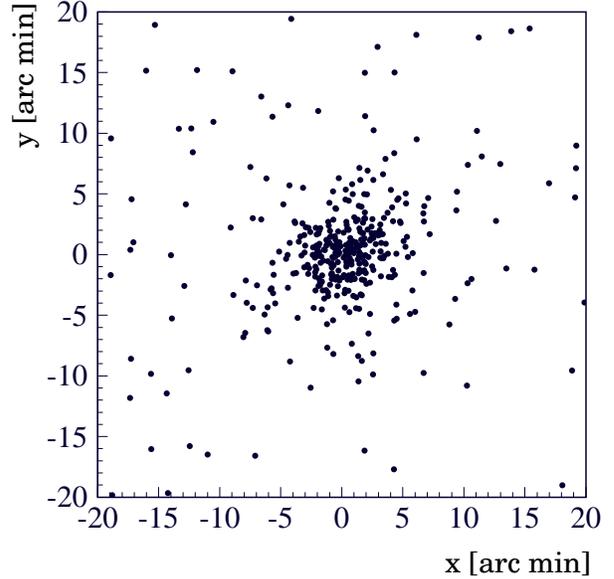}}
\label{fig_b}
\caption{Angular distribution of reconstructed 
showers in a local coordinate system centered 
on the Crab pulsar. Events were selected on the
basis of the predicted angular error ($<$ 3'
in both directions) and on the basis of image shapes
({\em mean scaled width} $< 1.2$).}
\end{center}
\end{figure}

While the evaluation of statistical errors is
straight forward, the control of systematic errors
is more difficult. The pointing of the telescopes is referenced to and
corrected offline on the basis of star images
(P\"uhlhofer et al. \cite{hegra_pointing}),
and the achievable pointing precision has been 
investigated in considerable detail. We are confident
to achieve a pointing deviation of less than $\pm 0.5'$
in each coordinate.
Indeed, when the Crab data set was subdivided into
seasonal subsets, the reconstructed source position
was in all cases consistent with the 
nominal source location -- the position of the 
Crab pulsar.

\begin{table*}[htb]
\begin{center}
\caption{Width of the angular distribution of events relative
to the source, comparing the Mrk 501 and Crab data sets with Monte Carlo
simulations using the measured gamma-ray energy spectrum as an input.
Data sets are selected according to the estimate of the angular
resolution, as provided by the shower reconstruction algorithm.
The quoted width values are derived using a Gaussian fit to the
projected angular distribution. For the last two rows of the table,
only the central part of the distribution is fit; there are
significant non-Gaussian tails (both in the Monte Carlo and in 
the data).}
\vspace{0.3cm}
\begin{tabular}{|c||c|c||c|c|}
\hline
Selection on & Mrk 501 MC & Mrk 501 data & Crab MC & Crab data \\
angular resolution [arc min]    & [arc min]   & [arc min] & [arc min]   & [arc min] \\
\hline
$\le$ 2.4 & $2.43 \pm 0.05$ & $2.46 \pm 0.06$ & $2.41 \pm 0.05$ & $2.41 \pm 0.14$ \\
$\le$ 3   & $2.81 \pm 0.04$ & $2.83 \pm 0.05$ & $2.81 \pm 0.04$ & $2.70 \pm 0.10$ \\
$\le$ 6   & $3.58 \pm 0.03$ & $3.63 \pm 0.04$ & $3.64 \pm 0.04$ & $3.70 \pm 0.09$ \\
all events & $4.23 \pm 0.04$ & $4.26 \pm 0.04$ & $4.30 \pm 0.04$ & $4.37 \pm 0.10$ \\
\hline
\end{tabular}
\label{taba}
\end{center}
\end{table*}

To evaluate the level at which the angular resolution
is understood, we use the sample of gamma-rays from the AGN Mrk 501 
as a reference set (see 
Aharonian et al. (\cite{hegra_501},
\cite{hegra_501b}) for details on this data set),
assuming that Mrk 501 represents a point source.
Table~\ref{taba}, columns 2,3 compare the measured angular distribution
for different subsets of events with the 
Monte-Carlo predictions. `Angular resolution' again refers
to the Gaussian width of the projected angular distribution 
of events. Excellent agreement between data and
Monte-Carlo is seen for all
data sets. We note that the resolution estimates given by the 
reconstruction algorithm are low by 10\% to 15\%, for the
samples selected for good resolution. The $\le$ 2.4' sample, e.g.,
should show a 2.2' resolution, compared to the measured value of
2.46'. Given the relatively crude parametrization of image parameter
errors used in the reconstruction (Hofmann et al., \cite{hegra_reco}), 
a deviation at this level is
not unexpected, and in any case the effect is fully reproduced by
the simulations. 

\begin{figure}
\begin{center}
\mbox{
\epsfxsize7.5cm
\epsffile{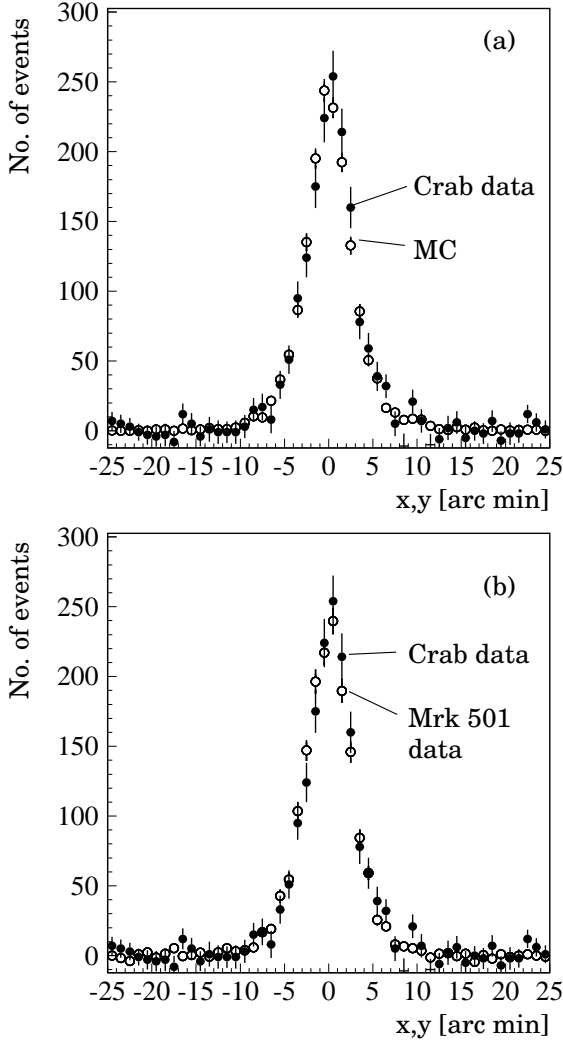}}
\label{fig_a}
\caption{(a) Full points: background-subtracted angular distribution 
of gamma-rays, projected onto the axes $(x,y)$ of a local coordinate
system centered on the Crab pulsar. Events are selected to provide an
angular resolution of better than 3'. Superimposed, as
open points, the corresponding distribution of Monte-Carlo events,
normalized to the same area. (b) Comparison of the angular distributions
of gamma rays from the Crab Nebula (full points) and from Mrk 501
(open points).}
\end{center}
\end{figure}

After these preliminaries, we can now address the Crab data set.
Fig. \ref{fig_a}(a) shows the background-subtracted
angular distribution of reconstructed showers relative
to the source direction, projected onto the axes of
a local coordinate system, after a cut on the estimated
error of less than 3'. Superimposed is the
corresponding distribution of Monte-Carlo events,
generated with the measured Crab spectrum. 
Fig. \ref{fig_a}(b) shows the corresponding
comparison with gamma-rays from Mrk 501. 
In particular after the selection on good angular resolution,
the two event samples are very similar in 
their characteristics (mean predicted resolution,
mean number of telescopes contributing to the reconstruction,
etc.) and can be compared directly, despite the differences
in the energy spectrum of the two sources. 
Table \ref{taba}, cols. 4,5 list the widths of the distributions
for the Crab gamma-rays, and the corresponding simulations.
In general, we find, within the statistical errors, good
agreement between the Crab and Mrk 501 data sets, and between
Crab data and Monte-Carlo.

Another option to
allow a direct comparison of the Mrk 501 and Crab data sets
is not to look at the 
angular deviations $(x,y)$ between a gamma-ray and the source,
but rather to normalize these quantities
with the angular resolutions $\sigma_{o,x}, \sigma_{o,y}$ 
predicted event-by-event
by the reconstruction algorithm. Ideally, one expects to see a
Gaussian distribution of unit width, independently of the energy
spectrum, the number of telescopes active for a given data set, etc.
The observed distributions are indeed Gaussian; as mentioned above,
their widths differ by up to 10\% to 15\% from unity, depending
on the selection of events. Most importantly, however, these width
values are consistent between all four sets of events (Mrk 501 MC,
Mrk 501 data, Crab MC, Crab data), within errors of 1.5\% or less for 
the large-statistics sets where all events are included. This agreement
demonstrates that there are no uncontrolled systematic effects
between the two experimental data sets, or between the experimental
data and the simulations.

Since the width of the angular distribution of gamma-rays 
from the Crab Nebula is consistent with the expected width,
and with the width observed for Mrk 501, we can only give
an upper limit on the source size. Taking into account the
statistical errors on the Crab sample and on the reference
samples, we find -- following Caso et al. (\cite{pdg}) -- 
99\% confidence level upper limits of
1.0' for the sample with a cut at 3' resolution,
and 1.3' for the $\le$ 2.4' sample. 
To be conservative, and since it was always planned to use the
$\le$ 2.4' sample as a safest compromise between statistical
and systematic uncertainties, we adopt the 1.3' limit.
Adding in additional systematic uncertainties due to
pointing precision, we quote a final limit of 1.5' for
the rms source size at a median energy of 2 TeV.

A possible contribution of gamma-rays from hadronic
processes is generally expected to be most relevant
at higher energies, in the 10 TeV to 100 TeV range. Therefore, the
source size was also studied for an event sample with
reconstructed energies above 5 TeV; data
are consistent with a point source and the 
corresponding limit on the rms source size is 
1.7'. Limited statistics prevent studies at even higher energies.

{The elliptical shape of the X-ray emission region suggests
to perform the same analysis in a rotated coordinate system,
with its axes aligned along the major and minor axes of the X-ray
profile. Results obtained for the width in the major and minor
direction do not show any significant difference. Given the fact that
the limits are large compared to the (rms-)size of the X-ray emission
region, this observation is not surprising. 
The TeV source is reconstructed 0.2' from the location of the 
pulsar, and is consistent both with the location of the pulsar,
and with the center of gravity of the X-ray emission region,
within the systematic errors
in the telescope
pointing (less than $\pm 0.5'$ in each coordinate). 
While the statistical error alone
is small enough to resolve a shift of 0.3' as observed in the X-ray image,
current systematic errors prevent such a measurement at TeV
energies.}

The limits obtained in this work are included in Fig. \ref{crab_size}.

{We note in passing that also the widths of the distributions
obtained for the AGN Mrk 501 are consistent with MC expectations,
indicating the absence of a halo on the arcminute scale. Potential
halo types for AGNs include wide pair halos (Aharonian, Coppi \& V\"olk 1994) 
or narrow halos caused
by intergalactic magnetic fields (Plaga 1995). For a relatively near source
such as Mrk 501, a pair halo would be much wider than the field of 
view of the camera and could not be detected as an increased apparent
source size; the other type of halo speculated in (Plaga 1995) would be well
below our resolution. We do not see any time-dependence of the source
size, or any correlation with the TeV gamma-ray flux.
Details will be given elsewhere.}

\section{Concluding remarks}

{The size limits given in this work
illustrate the precision which can nowadays be reached
in TeV gamma-ray astrophysics; a number of potential galactic and
extragalatic sources are predicted to be extended sources on this
scale.}

A few remarks concerning the interpretation
of the limits: the values quoted above refer to the rms source size,
with the implicit assumption that source strength is 
distributed over an area which is similar to, or small
compared to the angular resolution of the instrument, such
that the convolution of the source distribution and the 
Gaussian response function can again be approximated by a 
Gaussian distribution. This is obviously the case for a 
roughly Gaussian source, and was explicitly checked for
two alternative distributions, namely (a) the case that
the source strength is uniformly distributed over the 
surface of a sphere of radius $r$,
resulting in an rms source width of $r/\sqrt{3}$, and
(b) the case that the source is uniformly distributed
inside a sphere of radius $r$, with an rms size of $r/\sqrt{5}$.
The limit on 1.5' rms size can be reliably translated
into a limit on the radius of a surface source of 2.6',
or on the radius of a volume source of 3.4'. We note,
however, that one can construct source models with
a large rms source width which would not be detected by
our method. One example is a combination of a localized 
source at the center with a weak halo with an extension
of a few degrees or more. Such a faint, diffuse halo could
not be detected, yet formally results in a large rms width
of the source. However, given that it is hard to imagine
to have TeV emission even beyond the radio emission region,
such scenarios are hardly relevant for the Crab Nebula.

The limit on the size of the TeV emission region of the
Crab Nebula is, by a factor around 4, larger than the
size predicted by the standard inverse Compton models for gamma-ray
production in the nebula. 
The limits, however, approach the sizes expected for 
hadronic production models, where high-energy gamma-rays 
are produced by nucleon interactions,
more or less uniformly throughout the nebula.

\section*{Acknowledgments}

We profited from discussions with O.C. De Jager both on the
emission mechanism in the Crab Nebula, and on Crab data analysis. 
The support of the HEGRA experiment by the German Ministry for Research 
and Technology BMBF and by the Spanish Research Council
CYCIT is acknowledged. We are grateful to the Instituto
de Astrofisica de Canarias for the use of the site and
for providing excellent working conditions. We gratefully
acknowledge the technical support staff of Heidelberg,
Kiel, Munich, and Yerevan. GPR acknowledges
receipt of a Humboldt Foundation postdoctoral fellowship.

\end{document}